# Designing a π-based Programming Language in the .NET Framework: CLR Interoperability from the Programmer's Point of View

Manuel Mazzara

Department of Computer Science, University of Bologna, Italy

**Abstract.** Interoperability is the ability of a programming language to work together with systems based on different languages and paradigms. Presently, many widely used high-level languages implementations are providing access to external functionalities. Since we are researching for a new concurrent programming language design, our hope would be to see its widespread adoption in the next future. For this reason, such a language should include a way to be interfaced with already existing languages and older code. The aim of this paper is to rise briefly some ideas about the topic of CLR interoperability from the programmer's point of view. In particular, we want to focus our discussion on which kind of constructs a programmer would like to use for coping with these problems.

# 1 Toward a Foundation of Concurrency

The main motivation for designing a new concurrent programming language is that we need good programming abstractions for concurrency. Currently, the practice is to use sequential programming with the addition of some ad-hoc design principles (for example protecting every shared variable with a lock) and some concurrency supports in the language. Most popular programming languages treat concurrency not as a language feature but as a collection of external libraries which sometimes are not thoroughly specified. However, today more and more programmers are dealing with problems about concurrent and distributed programming. For this reason, to simplify design and development, concurrency and communication must necessarily be core language-level features and part of language specifications instead of add-ons or libraries. We strongly believe that, from the point of view of programmers, the $\pi$-calculus [1] represents a natural high level linguistic abstraction for concurrency.

For this reason, we are designing and implementing a programming language as based on a variant of the $\pi$-calculus in which explicit fusions are introduced as a very natural mechanism for addressing distribution [2]. Furthermore, our underlying theory allows the introduction of types for concurrency which promise great benefits in terms of program semantic analysis. Several works in literature show that in this framework checking for critical properties in distributed programming is possible at compile time. Behavioural type systems for Process Algebras, in fact, permit us to ensure processes to be deadlock-free or to obey a particular protocol [3].

An already existing practical application of this theory is represented by the XLANG Scheduler in Microsoft Biztalk Server 2000 [4], a recent tool used to integrate business systems. XLANG — the internal orchestration language of Biztalk — is explicitly built on a model from the $\pi$-calculus for a rigorous mathematical basis.

# 2 Language Environment

The language environment is an XML-based computing environment designed with concurrency and distribution as core computing primitives. As you can see in fig.1, we can represent the architecture as composed by two primary set of components. We call the first ones **Design Time Components** and the second ones **Run Time Components**.

The Compiler reads plain text files written in the textual language and generates an intermediate representation. These generated files are XML documents representing the **Execution Format** which is at the center of our world. The reason to syntactically capture it in XML is to enable the transferring of program text between different tools and to permit its manipulation. Execution of intermediate files is performed via the Virtual Machine.

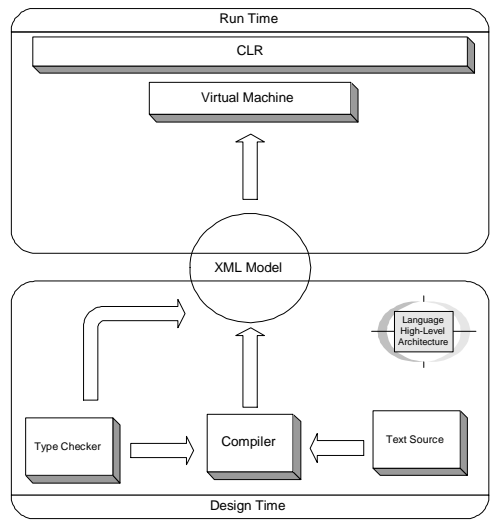

**Fig.1.** High Level Architecture

## 3 Interoperability

Interoperability is the ability of a programming language to work together with systems based on different languages and/or paradigms. Actually, we are researching for a new concurrent programming language design. If our hope is to see its widespread adoption in the next future, such a language should include a way for interfacing with already existing languages.

Microsoft .NET Common Language Runtime (CLR) is a platform which permits to interface different languages easily. It is intended to support a range of different languages and it provides a base for interoperation between them. In fact, code written for example in C# or VB can be compiled in the common Microsoft Intermediate Language (MSIL) [5]. In this context, it is easy to understand how a practical interfacing with the CLR can be suitable for integrating foreign functionalities This paper wants to introduce open issues in the language design related to CLR interoperability. Summarizing, this paper is about the following points:

- **Desired Feature**: Providing interoperability with existing languages.
- **Motivations**: We are proposing a new language. If our actual hope is to see its widespread adoption in the next future, such a language should include a way for interfacing it with already existing languages and older code.
- **Issues**: Many widely used high-level language implementations provide access to functionalities specified in a different language. An important characteristic of such mechanisms is the **interface** between our language and the other one.
- **Paper aim**: Individuate desired features and provide design options for CLR interoperability.

## 4 Channels for CLR Interoperability

The $\pi$-calculus is about communication. In effect, one relevant category of the language are channels. Channels are the way the processes have to communicate each other. What we address in this paper, however, is an issue of different nature. It concerns pragmatic. For our purposes, we need to classify channels in two different sets:

1. Channels for communication
2. Channels for CLR interoperability

The first category will be not addressed here. Channels for CLR interoperability, instead, are the focus of our discussion. In fact, we need to provide to this kind of channels information about how to instantiate the connection with the end-point and how to manage information flowing from and to it. A reasonable implementation could provide information to solve these points at the time in which the new statement for a particular channel is processed. At that time, the connection between the channel and his end-point should be instantiated. If the end-point is an object, the new statement could create an instance of it associating the channel and such an object. Messages sent to this channel have to be automatically translated into method calls and return values must be transmitted back synchronously. Anyway, we should pay particular attention here because a very subtle problem could arise.

Consider a scenario in which class constructors have side effects. Clearly, this is a bad programming scenario but bad programming scenarios are real life scenarios. If we bind the creation of a new instance of an object with the relative new port definition, we can violate the classical structural congruence rule:

$$(newx)(newy)P \equiv (newy)(newx)P$$

Hopefully this shouldn't happen, for this reason we want to separate the new semantic and the object instantiation.

## 5 Programming Constructs for CLR Interoperability

An interface for accessing foreign functionalities should be **compact** and clear for the programmer as is, for example, a class in OOP. In this way, a programmer can quickly understand the code meaning. Suppose to have the following C# toy class:

```
public class Account { public static readonly int Pin
    = 1976528; public static void readn(){ return;
     }
     public static int read(){
         return Pin;
     }
}
```

A syntactical interfacing block should be declared in our program to capture this binding. Here we propose the following but is just an example:

```
extern FClass -> class Account {
    void readn(){ call readn:
    void; return Ret1: void;
    } acceded as {rec S {readn().Ret1().S}} int
    read(){ call read: void; return Ret2: int;
    } acceded as {rec S {read().Ret2(int).S}} }
```

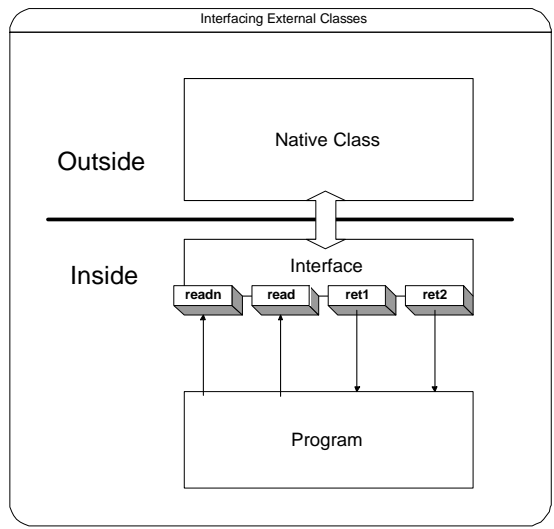

**Fig.2.** Components involved in accessing external functionalities

The exported channels list readn, read, Ret1, Ret2 has a global visibility and the interactive behavior of the class is defined after the keywords acceded as. This specification of the interactive behavior is about behavioral type systems as announced previously. A main point of our theoretical approach, in fact, is that it can lead to a compiler where the type system can check not just data types matching but also that the use of some channels (to send and receive data) is obeying a given protocol with these messages (and with these channels). Anyway, in this case it seems reasonable to allow only a basic interaction schema for a single method, i.e. to allow a call followed by a return recursively and nothing else. Different behaviors seem very weird to be allowed. If we decide to accept this claim, the keyword acceded as and what follows can be eliminated.

## 6 The Programmer's Point of View

Approaching the problem in this way, we are defining an interface for an external class which is really compact and where all the involved pieces are grouped together, so it can be easy reading and understanding what the code is actually doing. In fact, channels are first citizens of the language and this solution focuses the interface on them. It is easy to locate at a first glance the access points for the class at the programmer level. Furthermore, because this interface is isomorphic to the external class, any future change in the external class can be very easily mapped inside the code and this can simplify both design an maintenance. Fig.2 represents the involved components, as you can see the interface is a *collection of channels* and these channels are the only access points for the class. We have one port for each accessible method. This encapsulation is close to OOP concept,
i.e. an average programmer is already familiar with it. Another advantage of this approach is that we cannot affect the new semantics as showed above because a new instance of an object is not generated at the time in which the statement is processed.
Furthermore, the process of interfacing a program for accessing foreign functionalities is very easily automatable. In fact, a tool which loads an MSIL assembly file and scans each method extracting the information necessary to build an interface like the one showed above is quickly implementable. This allows to use external code written in a range of different languages supported by the .NET framework.

## 7 Conclusions

In this paper, we introduced the problem of CLR interoperability for a $\pi$-calculus based programming language we are designing and developing. After a brief overview of the main issues concerning such a language, we arose some ideas about how to interface foreign functionalities. A key point of our discussion have been channels; since the $\pi$-calculus is about communication they are a relevant category of the language. As we said channels are the way the processes have to communicate each other. For our purposes, we pragmatically divided them in two different categories: channels for communication and channels for CLR interoperability. The second category is the way programmers have to access, through the MSIL, foreign functionalities written in different languages. Programmers need reasonable constructs to access this external word, so we introduced some desirable features these constructs should have. We also showed a design option to solve the problem. Obviously, this work is not exhaustive and many other issues need to be arisen and developed in the future. Topics which deserve to be treated extensively are, for example, the concrete implementation of this kind of interoperability, the developing of tools for automatic generation of interfaces and many other low level subjects not addressed here.